# LOW TEMPERATURE THERMAL EXPANSION OF PURE AND INERT GAS-DOPED FULLERITE $C_{60}$.

**A.N. Aleksandrovskii[1], A.S Bakai[2], A.V. Dolbin[1], V.B. Esel'son[1], G.E. Gadd[3], V.G. Gavrilko[1], V.G. Manzhelii[1], S. Moricca[3], B. Sundqvist[4], B.G. Udovidchenko[1]**

[1] Institute for Low Temperature Physics & Engineering NASU, Kharkov 61103, Ukraine
[2] National Science Center "Kharkiv Institute of Physics & Technology", Kharkov 61103, Ukraine
[3] Australian Nuclear Science & Technology Organisation, NSW 2234, Australia
[4] Department of Experimental Physics, Umea University, SE - 901 87 Umea, Sweden

Electronic address: aleksandrovskii@ilt.kharkov.ua

**Abstract.**
The low temperature (2-24 K) thermal expansion of pure (single crystal and polycrystalline) $C_{60}$ and polycrystalline $C_{60}$ intercalated with He, Ne, Ar, and Kr has been investigated using high-resolution capacitance dilatometer. The investigation of the time dependence of the sample length variations $\Delta L(t)$ on heating by $\Delta T$ shows that the thermal expansion is determined by the sum of positive and negative contributions, which have different relaxation times.

The negative thermal expansion usually prevails at helium temperatures. The positive expansion is connected with the phonon thermalization of the system. The negative expansion is caused by reorientation of the $C_{60}$ molecules. It is assumed that the reorientation is of quantum character. The inert gas impurities affect very strongly the reorientation of the $C_{60}$ molecules especially at liquid helium temperatures.

A temperature hysteresis of the thermal expansion coefficient of Kr– and He-$C_{60}$ solutions has been revealed. The hysteresis is attributed to orientational polyamorphous transformation in these systems.

## 1. Introduction

Here we report the results obtained in a series of investigations of the low temperature (2-24 K) thermal expansion of pure $C_{60}$ and $C_{60}$ doped with inert gases.

In the first studies of the series [1,2] we detected a negative and very high in magnitude coefficient of the linear thermal expansion of polycrystalline and single crystal $C_{60}$ in the region of liquid helium temperatures. This unusual effect has stimulated our further research.

One of the characteristic features of fullerite $C_{60}$ is the essential influence of the orientational states of its molecules on the physical properties of the crystal. A molecule of fullerene $C_{60}$ is shaped as a truncated icosahedron whose surface forms 20 hexagons and 12 pentagons. The noncentral interaction between the globular molecules of $C_{60}$ is much weaker than the central one and fullerene molecules show reorientational motion over a wide temperature range. At condensation, fullerite $C_{60}$ forms a face-centered-cubic (fcc) lattice. As shown by experimental studies (see, for example, Ref. 3), the rotation of molecules in this phase is weakly hindered. On decreasing the temperature to about 260 K, a structure-orientational phase transition occurs into a low-temperature phase of Pa3 symmetry. This transition is accompanied by partial orientational ordering of the rotation axes of the $C_{60}$ molecules. The almost free rotation of molecules changes into rotation around the space diagonal <111>. On a further decrease in the temperature in the Pa3 phase, the rotational motion of the molecules around the <111> axes slows down. Near $T \approx 90$ K the rotation of molecules around the <111> axes is hindered almost entirely, and an orientational glass is formed.

Initially, our goal was to clear up the nature of the negative thermal expansion of $C_{60}$ In this context it was interesting and important to find out how impurities could influence this effect. It seemed natural to start with inert gases as impurities, which being comparatively simple atoms would hopefully facilitate the interpretation of the results. By introducing different inert gases (He, Ne, Ar, Kr) into the fullerite lattice, we intended to find how the sizes of the impurity atoms may influence the negative thermal expansion.

In the course of these investigations of $C_{60}$ doped with inert gases, new effects have been revealed, which appear worthy of special and independent consideration.

## 2. Experimental aspects

The linear thermal expansion coefficient was measured using a capacitive dilatometer [4] with a resolution $2\times10^{-9}$ cm, specially modified for measuring fullerite samples. The dilatometer was constructed so that all elements capable of affecting the measured results on account of their own thermal expansion, were in a bath of liquid helium at constant temperature. The scheme of the measuring cell of the dilatometer, the procedure for mounting the sample in the dilatometer and the measurement procedure have been described elsewhere [1,2]. The thermal expansion of fullerite $C_{60}$ was measured by a step change technique, described as follows. With the sample at a constant temperature $T_1$, the temperature of the objective table (containing the sample) was then changed to a temperature $T_2$ which from this moment was kept constant. Changes in temperature and sample length were registered once a minute and processed by a computer in real time. When the temperature drift of the sample did not exceed 0.01 K in 10 minutes, we determined the change of its length due to the change in temperature from $T_1$ to $T_2$. During the measurements the steps from $T_1$ to $T_2$ were 0.1 – 1 K depending on the temperature range.

The change in the sample length was determined during both a temperature increase and a decrease. The linear thermal expansion coefficient α was obtained by differentiation with respect to T of the temperature dependence of the relative elongation ΔL/L of the samples.

The low temperature thermal expansion of $C_{60}$ fullerite should be isotropic because it has a cubic lattice. Thus, in principle, both single crystal and polycrystalline $C_{60}$ can be used in studies of the thermal expansion.

Our polycrystalline samples were produced by compacting $C_{60}$ powder under a pressure 0.1 GPa. The stresses thus induced could affect the results of the subsequent dilatometric measurement. The gas molecules adsorbed at grain boundaries can also influence the results. In order to ascertain the presence of such effects, we also investigated $C_{60}$ in the form of a single crystal. To prepare the inert gas- $C_{60}$ sample, the single crystal is not subjected to the compression and mechanical treatment as in the polycrystal samples.

Four samples of pure $C_{60}$ were used – three polycrystals ($C_{60}$(I), $C_{60}$(III), $C_{60}$(IV)) and one single crystal ($C_{60}$(II)). In the subsequent experiments the samples $C_{60}$(III) and $C_{60}$(IV) were saturated with inert gasses. Information pertaining to the pure $C_{60}$ samples is given below.

**$C_{60}$ (I).** The sublimated $C_{60}$ powder for sample preparation was supplied by Term USA, Berkeley, CA, and had a nominal purity of better than 99.98%. No traces of solvents were found by Raman analysis within its accuracy (0.1% by mass). Room temperature powder X-ray diffraction pattern of the material displayed sharp peaks from fcc structure (a =14.13 A). In an atmosphere of dry argon the $C_{60}$ powder was loaded in a small piston-cylinder device used for the sample preparation. After subsequent compacting of the powder at about 0.1 GPa, the sample (pellet of 6 mm in diameter and about 2.4 mm in height, the grain sizes of 0.1-0.3 mm) was immediately transferred into a glass tube and dried under dynamic vacuum $10^{-6}$ Torr for about 16 hours. The compacting procedure was done in air and did not exceed 15 minutes. Finally, the sample was sealed in vacuum $10^{-6}$ shielded from light and kept in that state for 3 months until the beginning of dilatometric measurements.

**$C_{60}$ (II).** A very large $C_{60}$ single crystal ($\approx 6.5 \times 4.3 \times 3.1$ mm$^3$) was obtained from Dr. M. Haluska, Vienna. This crystal was grown by the sublimation method in an ampoule under vacuum from Hoechst "Super Gold Grade" $C_{60}$. The crystal was never subjected to air or oxygen after growing was completed, and was transferred from its growth tube whilst under Ar, into a glass ampoule, which was then evacuated before sealing.

**$C_{60}$ (III).** The sample was a cylinder 9 mm high and 10 mm in diameter. The sample characteristics and the preparation condition are similar to those for the sample $C_{60}$(I).

**$C_{60}$ (IV).** The $C_{60}$ powder was supplied by "SES research" company and had a nominal purity 99.99%. The condition of compacting the sample $C_{60}$(IV) was similar to that for the sample $C_{60}$(I). The sample was 6.5 mm in height and 10 mm in diameter. The average grain size was 0.1 mm. The powder and the sample prepared were kept in the air. Immediately prior to measurement of the thermal expansion, the sample was evacuated dynamically for six days.

It is important to note that all of our fullerite samples had a purity better than 99.98%.

The procedure for mounting the samples in the dilatometer were as follows. Prior to sample mounting, the glass ampoule containing the sample was opened in an argon atmosphere at with an excess pressure around 200 Torr. During the process of sample mounting into the measuring cell of the dilatometer, the sample was exposed to the air for no more than 20 minutes and then was evacuated. During the measurements the vacuum in the dilatometer cell was maintained at the level of $10^{-6}$ Torr. The measurement procedure was the same for all samples.

## 3. Results and discussion
### 3.1 Pure $C_{60}$

The typical time dependence of the change in the $C_{60}$ sample length $\Delta L$ after a fast increase in the temperature of the objective table (with the sample) by $\Delta T$ is shown in Fig.1. At each temperature the sign of the total thermal expansion is determined by two competing mechanisms responsible for expansion and contraction of the sample.

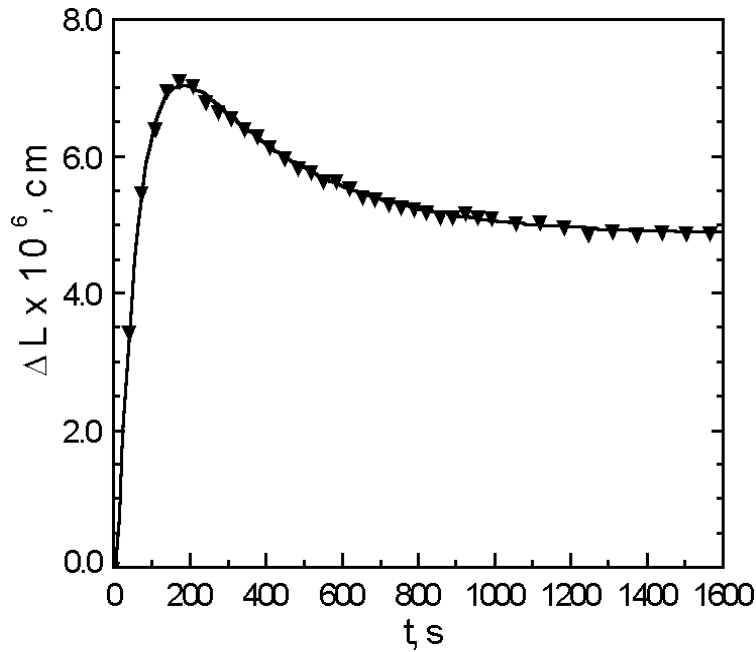

**Fig. 1.** The characteristic time dependence of the sample length variations $\Delta L(t)$ on heating by $\Delta T$

The temperature dependence of the thermal expansion coefficient for the four pure $C_{60}$ samples is shown in Fig.2. In the same figure the open circles show the linear expansion coefficients measured dilatometrically by Gugenberger et. al. [5] on single crystal $C_{60}$. Note that until the studies reported here, the thermal expansion of $C_{60}$ has not been investigated below 5K.

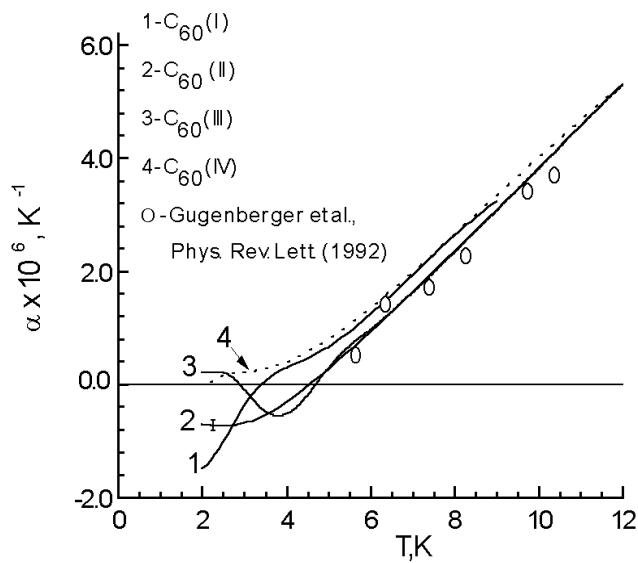

**Fig.2.** Temperature dependence of the thermal expansion coefficient of pure fullerite $C_{60}$.

Above 7K the results obtained are all in good agreement. Considerable discordance is however observed for the different samples in the region of liquid helium temperatures. In this region the thermal expansion coefficient for samples I, II and III have unusually high negative values for such low temperatures, whilst that of sample IV, has in contrast always has a positive value. The discrepancy in the α-values for different samples will be discussed in the subsequent section. Here we shall attempt to explain the nontrivial effect of negative thermal expansion of $C_{60}$.

A possible mechanism responsible for the negative thermal expansion of fullerite $C_{60}$ might be rotational tunnelling of molecules. Sheard [6] was the first who paid attention to the fact that rotational tunnelling of molecules can lead to a negative thermal expansion. This problem was considered in detail by Freiman [7] as applied to the thermal expansion of solid methane.

The large moment of inertia of the $C_{60}$ molecule is a strong objection to the assumption of a rotational tunneling mechanism as producing or the negative thermal expansion coefficient for $C_{60}$ fullerite below liquid helium temperatures. In the orientational glass, which evolves in the fullerite below 90K, the potential barriers $U_\phi$ impeding rotation of the molecules can vary within wide limits. The tunnel rotation can be performed only by those molecules for which the $U_\phi$- barriers are quite low. $C_{60}$ molecules that experience such barriers are referred to as defects. If the $U_\phi$-barriers grow with a decrease in the crystal volume, the tunnel rotation leads to a negative thermal expansion of the crystal [6,7].

In the case of tunnel rotation the absolute values of the Gruneisen coefficient can be very high [2,8]. In Fig 3 [2] the temperature dependence of the Gruneisen coefficient is shown for the samples $C_{60}(I)$ and $C_{60}(II)$.

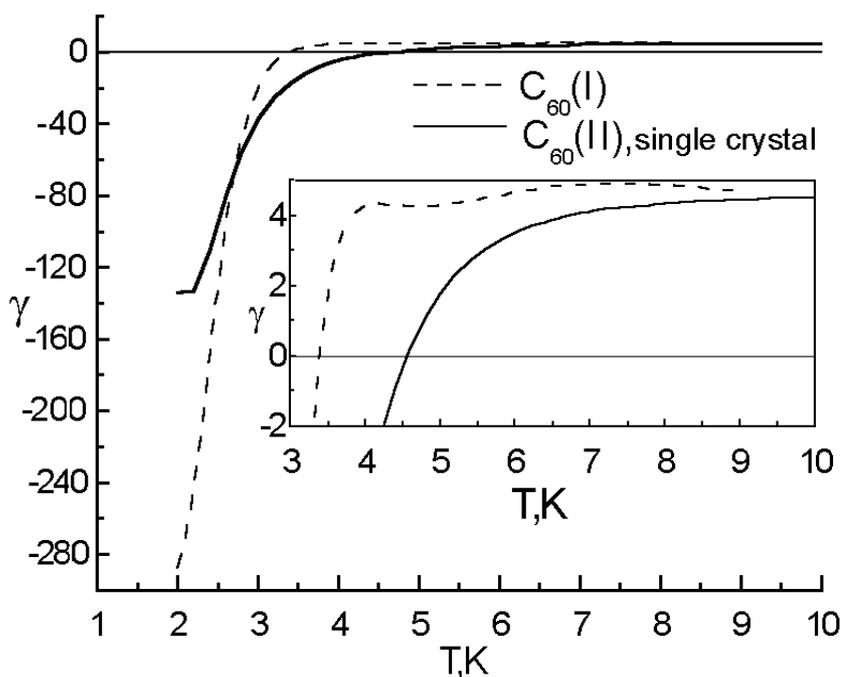

**Fig.3.** Gruneisen parameter of fullerite $C_{60}$

It should be noted here that the Gruneisen coefficients for the phonon and libron spectra of molecular crystals are usually of the order of one [9]. The unusually large negative values of γ testify in favour of the assumption of the tunnelling nature of the negative thermal expansion of fullerite $C_{60}$.

The factors that might cause the negative thermal expansion of fullerite $C_{60}$ were analyzed in [10, 11]. Since the $C_{60}$ molecules have five- fold symmetry axes, they cannot entirely be orientationally ordered, hence, certain defects are inevitable at low temperatures. The negative thermal expansion was explained qualitatively [11] assuming co-existence of different orientational domains in the simple cubic lattice of $C_{60}$. The potential relief is much smoother in the inter-domain space (domain walls) than inside the domains and the $U_\phi$-barriers impeding rotation of the molecules are much lower in the domain walls. As a result, the molecules in the domain walls (unlike those inside the domains) can remain nearly free rotors down to much lower temperatures.

**3.2 $C_{60}$ doped with Ar and Ne.**

We investigated the thermal expansion coefficient α of the $C_{60}$(III) sample doped with Ar and Ne. [12,13]. The penetration of gas atoms into the fullerite lattice was expected to change the $U_\phi$-barriers and hence effect rotation of the neighboring $C_{60}$ molecules. The impurity atoms could thus influence the probability of rotational tunneling of these molecules and the tunneling-induced contribution to the thermal expansion. If the $U_\phi$-barrier height and (or) width increase, the total negative thermal expansion $\int \alpha dT$ should decrease and shift towards lower temperatures. Recall that in the low temperature phase each $C_{60}$ molecule is associated with two tetrahedral and one octahedral interstitial cavities whose average linear dimensions are about 2.2 Å and 4.2 Å, respectively [14,15]. According to X-ray and neutron diffraction data [16-19], the Ne and Ar atoms with the gas-kinetic diameters 2.788 Å and 3.405 Å [20], respectively, occupy only the octahedral cavities.

The $C_{60}$(III) sample was doped with neon and argon at room temperature under atmospheric pressure. The doping lasted for 340 hours for neon and 460 hours for argon. In these experiments we did not measure the Ne and Ar concentrations in fullerite. According to [16], on $C_{60}$ saturation with Ne at room temperature under atmospheric pressure, the Ne concentration in $C_{60}$ reaches 20 mol%.

Figure 4 shows the measured thermal expansion coefficients of fullerite before doping (curve 1) and after doping with neon (curve 2) and argon (curve 3). It is seen that the doping has strongly affected the thermal expansion of $C_{60}$: the introduction of the doping gas reduced considerably the positive coefficients, suppressed strongly the effect of negative thermal expansion and shifted it towards lower temperatures. To explain the effects observed, it seems natural to assume that the atoms of impurities impede rotation of the $C_{60}$ molecules and thus enhance the noncentral forces acting upon the $C_{60}$ molecules. With the noncentral forces enhanced, the libration frequencies $\omega_i$ of the $C_{60}$ molecules should increase and at $T < \hbar\omega_i/k$ the contribution of librations excitations to the heat capacity and thermal expansion of the crystal should decrease. At T>6 K the tunneling effects are no longer important. In this temperature range the translational lattice vibrations and the molecule librations are responsible for the thermal expansion. In [21,22] the libration excitation contribution to the heat capacity of $C_{60}$ was described using two librations frequencies with energies equivalent to 30 K and 58 K. Thus, the inequality $T<h\omega_i/k$ holds for the temperature interval 5-12 K and the assumption of a decrease in the libration excitation contribution to the thermal expansion at these temperatures seems to be reasonable. As the noncentral forces increase, the $U_\phi$- barriers impeding rotation of the molecules become higher. This diminishes the probability of rotational $C_{60}$ tunneling and the tunneling splitting of the energy levels. As a result, the negative thermal expansion decreases and the region of tunnel effects shifts towards lower temperatures.

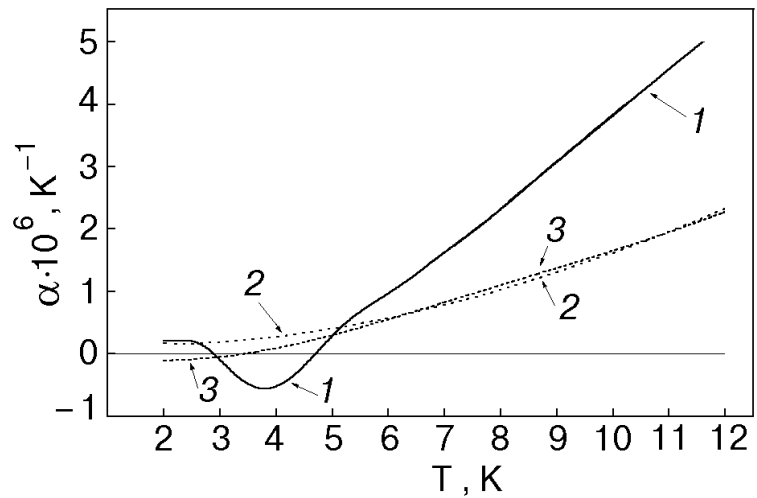

**Fig.4.** Linear thermal expansion coefficients: pure fullerite (1), fullerite doped with neon (2), fullerite doped with argon (3).

We should remember that the libration excitation contribution is made by all $C_{60}$ molecules, while the negative contribution is made only by the molecules which we call "defects".

An interesting feature was exhibited by the linear thermal expansion coefficient of Ne- and Ar-doped $C_{60}$ (see fig.4). At 6-12 K the $\alpha(T)$ values practically coincide for Ar-and Ne-saturated samples. Let us assume that doping with Ar and Ne increases the libration frequencies of the $C_{60}$ molecules to such extent that the contribution of librations excitations to the thermal expansion is negligible up to 12 K. Besides, it is natural to expect that the Ar and Ne impurities have negligible effect on the translational vibration of the lattice and their contribution to the thermal expansion. If this is correct, the thermal expansions of the Ne-$C_{60}$ and Ar-$C_{60}$ solutions and the contribution of the translational vibrations to the thermal expansion of pure $C_{60}$ should coincide in the 6-12K region where the contributions of librations excitations and tunneling to $\alpha(T)$ are not observable. We tried to describe the contribution of translational vibrations as

$$\Delta\alpha_{trans} = \gamma C(T/\Theta_D)/BV \qquad (1)$$

where $\gamma$ is the Gruneisen coefficient, V is the molar volume of $C_{60}$, B is the bulk modulus of $C_{60}$, $C(T/\Theta_D)$ is the Debye heat capacity, $\Theta_D$ is the Debye temperature of $C_{60}$. The calculation was made using $V = 416.7$ cm$^3$/mol [23] and $B = 10.3$ GPa [24].

$\gamma$ and $\Theta_D$ were fitting parameters. $\Delta\alpha_{trans}$ can be described well by the above expression (1) taking $\gamma = 2.68$ and $\Theta_D = 54$ K. $\gamma$ is close to the corresponding values for rare gas solids [20]. Literature data for $\Theta_D$ of $C_{60}$ are much scattered. The analysis of the temperature - heat capacity dependences yields $\Theta_D$ =37 K [22], 50 K [25, 26], 60 K [27], 80 K [28] and 188 K [29]. $\Theta_D$ =100 K was obtained in studies of elastic properties on single crystal $C_{60}$ [30]. Even if we disregard too high $\Theta_0$ =188 K [29] and $\Theta_D$ =100 K, where the error is quite appreciable [30], the scatter of $\Theta_D$ values is still wide. The Debye temperatures obtained by the calorimetric method are highly sensitive to the impurities present in fullerite and the technique employed to separate the contributions to heat capacity. The $\Theta_D$ results calculated from ultrasound velocities are more reliable since they are mainly determined by the translational vibrations of the lattice and are low-sensitive to impurities. The ultrasound data for single crystal $C_{60}$ at T=300 K provide $\Theta_D = 66$ K [31]. In [23] literature data on ultrasound velocities of polycrystalline $C_{60}$ were analyzed and extrapolated to low temperatures. The Debye temperature thus calculated at T = 0 is 55.4 K, which is very close to our result.

The above consideration suggests that our $\gamma$ and $\Theta_D$ are quite realistic and thus strengthens the assumption that curves 2 and 3 in Fig.4 describe the contribution of translational vibrations to the thermal expansion coefficient of pure $C_{60}$. Proceeding from this assumption, we can take the difference between curve 1 and curves 2, 3 (Fig.4) at 6-12 K as the contribution of librations excitations $\Delta\alpha_{lib}$ to the thermal expansion coefficient of pure $C_{60}$. $\Delta\alpha_{lib}$ is well described by the Einstein term with the characteristic Einstein temperature $\Theta_E$=39 K. This value does not come into conflict with the above results [22]. Note that the idea of the impeding effect of the impurity gas molecules upon the rotational motion of the $C_{60}$ molecules was put forward earlier in [32]. The idea is supported by the fact that at low temperatures (T=15 K) the lattice parameter of a saturated $Ar_xC_{60}$ solution is 0.006 Å smaller than that of pure fullerite [18].

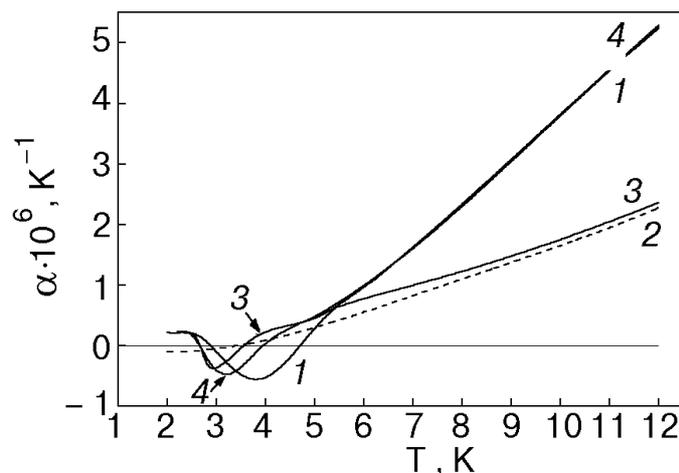

**Fig.5.** Temperature dependence of the thermal expansion coefficients: pure fullerite (1), Ar-doped fullerite (2); fullerite after evacuation of Ar for 3 days (3) and for 45 days (4).

To obtain more information, we studied how the thermal expansion coefficient changed when the doping atoms were removed from the sample. For this purpose, the measuring cell with the sample was warmed to room temperature and evacuated to $1\cdot 10^{-3}$ Torr. The gas evacuation at room temperature lasted for 3 days. The thermal expansion was then measured at low temperatures. The results are shown in Fig.5 (curve 3). It is seen that the thermal expansion coefficient changes only slightly above 5 K, but below 3.5 K the negative thermal expansion now again has the minimum typical for undoped $C_{60}$. The measuring cell with the sample was warmed again to room temperature and gas evacuation was continued for 42 days. The thermal expansion coefficients measured there after are shown in Fig. 5 (curve 4). Note, that after a total of 45 days evacuation of argon the "high-temperature" part of the thermal expansion coefficient was restored completely. The negative thermal expansion in the range 2.5-5 K was, however, still different from the value for the initial pure sample. Similar experiments were made on a sample saturated with neon. After a 45 days' exposure of the sample to room temperature and vacuum, the results obtained before and after doping coincided in the whole temperature interval. We can thus conclude that the $C_{60}$ sample was completely free of neon. The desaturation of this duration was not sufficient to remove all argon from the sample.

The changes in $\alpha(T)$ after removal of argon from the sample may be explained as follows. The molecules, we called "defects", in Sect. 3.1 become permanently displaced with respect to the lattice sites. As a result, the octahedral voids surrounding such molecules are not identical, and the potential wells that they form for the impurity are different too – deeper or shallower than the octahedral potential wells adjacent to most of the $C_{60}$ molecules. At the initial stage of evacuation, the Ar atoms first leave the shallow potential wells near the defects, and this causes faster changes in the negative part of $\alpha(T)$. The removal of the Ar atoms from the deeper wells proceeds slower than in most of the crystal volume. That is why at the completing stage, the positive part of $\alpha(T)$ is restored faster at T > 5 K.

This consideration leads us to assume that the distinctions in the behavior of $\alpha(T)$ for different pure $C_{60}$ samples (see Fig.2) can be attributed to different amounts of the "residual" impurity whose particles are still present in the deep potential wells near the defects. The negative thermal expansion should therefore be particularly sensitive to the pre-history of the sample.

We believe that the absence of the negative contribution to the thermal expansion of pure $C_{60}$ (sample IV) down to 2 K is connected with saturation of the sample with air gases. The results of the TGA analysis also evidence that appreciable amounts of gas are present in sample.

### 3.3. $C_{60}$ doped with Kr and He.

It was noted above that on doping with Ar the molar volume of $C_{60}$ decreased at T<60K [18]. Like Ar atoms, the introduced Ne atoms occupy only octahedral interstitials, but their gas-kinetic diameter is smaller. It is natural to assume that at low temperatures the doping with Ne also leads to a decrease in the average distance between the $C_{60}$ molecules in the lattice. The above discussion agrees with the conclusion that Ar and Ne impurities impede the rotational motion of $C_{60}$ molecules.

Our next goal was to investigate how the impurities increasing the volume of fullerite affect its thermal expansion. The impurities chosen were Kr and He. When admixed, the Kr atoms occupy the octahedral interstitial cavities and because of rather large gas-kinetic diameter of Kr atoms the volume of the solid $C_{60}$ increases [18]. When He impurity is added [33], the volume of $C_{60}$ increases presumably because the He atoms occupy not only the octahedral interstitial cavities but the much smaller tetrahedral ones as well [32, 33].

**3.3.1. $C_{60}$-Kr.** The $C_{60}$(IV) powder was saturated with Kr at T=500°C and a Kr pressure of 170-200 MPa for 43 hours. The Kr concentration was found from the thermal gravimetric analysis [18] to be 60 mole %. The powder compacting, the procedure of mounting the sample in the dilatometer and the measurement procedure were similar to those applied to $C_{60}$(I).

The thermal expansion of $C_{60}$ doped with Kr was measured in six experimental runs. The temperature dependence of the thermal expansion coefficient α(T) is illustrated in Fig.6. The measurement was made on heating (solid circles and triangles) and cooling (open squares and the smoothed solid curve) the sample. Note two important features in the behaviour of the thermal expansion coefficient.

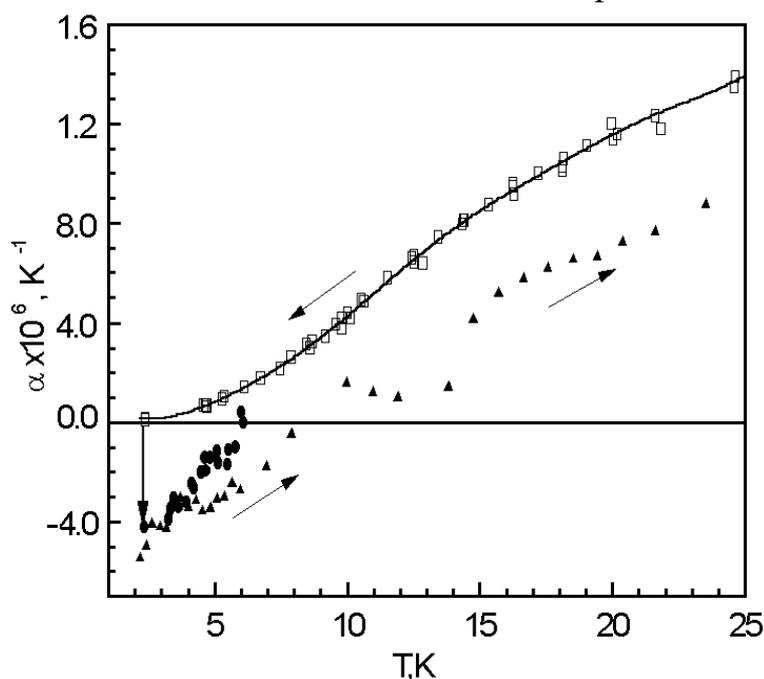

**Fig.6.** Temperature dependence of the thermal expansion coefficients of Kr-doped fullerite

First, there is a hysteresis of the temperature dependence α(T). Let us recall that α(T) hysteresis was not observed for pure $C_{60}$ and $C_{60}$ doped with small amounts of Ne and Ar. The results obtained in different measurement runs on the sample being cooled down from the highest measurement temperature (24 K) exhibit good agreement. The deviation from the smoothed curve is not large. The agreement is however much worse for the sample being heated from the liquid helium temperatures. The scatter in the data is quite appreciable even within one measurement run. To avoid extra complexity of Fig. 6, the results of only two measurements made on heating the sample are shown in the figure. We however emphasize that in all the cases the α-values measured on cooling are higher than α obtained on heating and exceed the thermal expansion coefficient of the initial $C_{60}$(IV) sample at T>10 K. The excess increases with temperature and reaches 15% at T=22 K. Another important point should be noted. To change from α measured on cooling to α measured on heating at the lowest experimental temperature (2 K), we had to keep the sample at T≈2 K no less than for three hours. As an example, we detail the procedure. On cooling the sample down from 24 K we reach T=2.4 K and start measuring the thermal expansion on heating. The values of the first two measurements are close to the results of extrapolation from the α(T) dependence for the cooling case following α(T) ~ $T^3$. On the subsequent measurement the thermal expansion coefficient changed abruptly to the value indicated by the vertical arrow in Fig.6.

Now we discuss the other feature in the temperature dependence of the thermal expansion coefficient for the Kr-$C_{60}$(IV) solution. The α – values obtained on heating the sample in the region of liquid helium temperatures become negative. Their absolute values are several times higher than those for the pure $C_{60}$(I), $C_{60}$(II), $C_{60}$(III) samples and for $C_{60}$(III) doped with Ne and Ar. It is interesting that the $C_{60}$(IV) sample itself (we used it for doping with Kr) has no negative thermal expansion in the whole temperature interval (2-24 K).

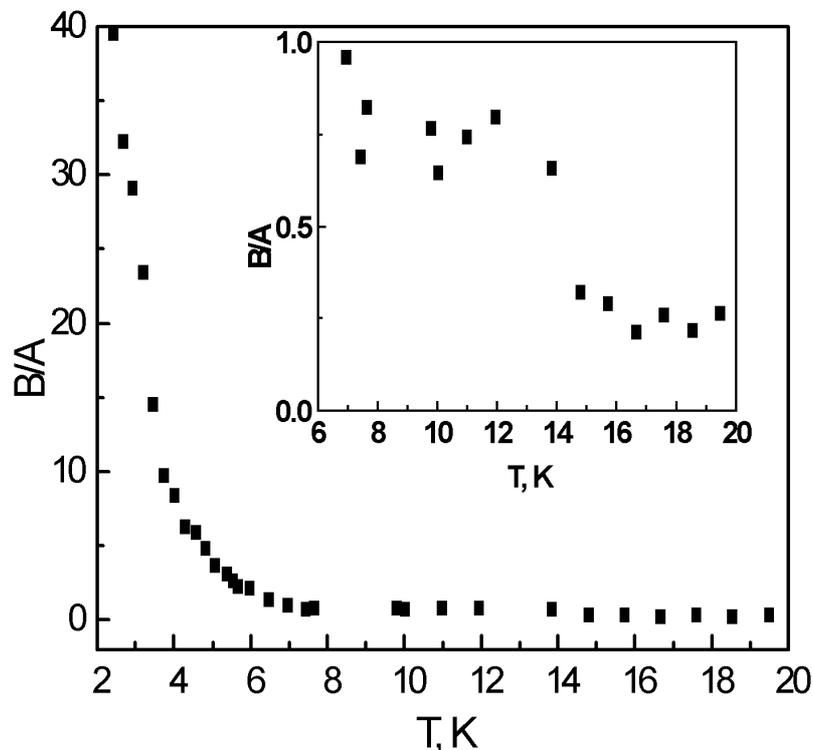

**Fig.7.** The relation between negative and positive contributions to the thermal expansion of Kr-doped $C_{60}$.

No negative contribution to the thermal expansion was detected on cooling the $C_{60}$(IV)+Kr sample from the highest temperature (24 K) to 2 K. However, if the sample was previously kept at T = 2 K over three hours and then cycled in the interval 2-5 K, no α(T) hysteresis is observed. The negative contribution to the thermal expansion exists both on cooling and on heating. Let us analyze the behavior of the negative contribution to the thermal expansion of the sample. The characteristic time dependence of the sample length variations ΔL(t) on heating by ΔT is shown in Fig.1. We describe this dependence as:

$$\Delta L(t) = A(1 - \exp(-t/\tau_1)) + B(\exp(-t/\tau_2) - 1) \qquad (2)$$

where the first term describes the positive contribution and the second term stands for the negative one; A and B are the absolute values of the corresponding contributions at $t \to \infty$; $\tau_1$ and $\tau_2$ are the characteristic relaxation times for these contributions.

The B/A value is the ratio between the negative and positive contributions to thermal expansion. The averaged temperature dependence of the B/A ratio calculated from the data obtained on heating the sample is shown in Fig.7. The negative contribution prevails at liquid helium temperatures and remains significant up to 20K.

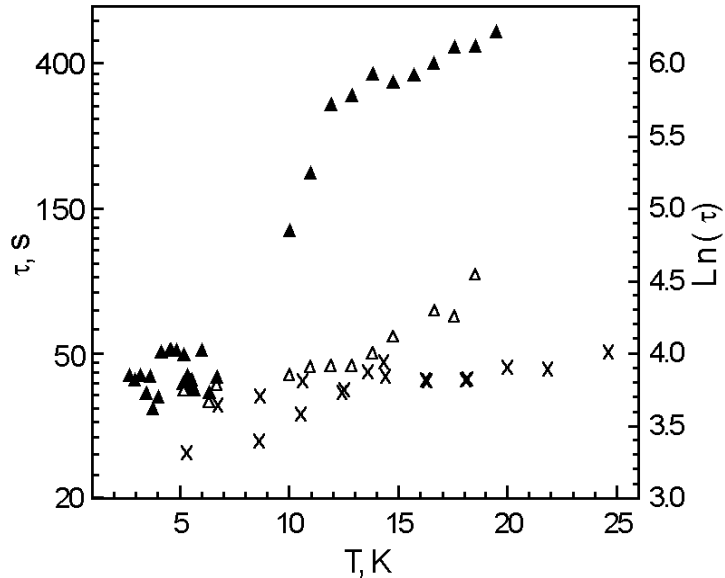

**Fig.8.** Characteristic relaxation times of positive $\tau_1$(Δ, ×) and negative $\tau_2$(▲) contributions to the thermal expansion of Kr-doped $C_{60}$.

The temperature dependences of the relaxation times $\tau_1$ and $\tau_2$ are illustrated in Fig.8. The $\tau_1$ values were calculated on both heating (Δ) and cooling (×) the sample.

We propose the following explanation of the features observed in the thermal expansion of the Kr-$C_{60}$ solution. The hysteresis and the jump-like change in the thermal expansion coefficient of $C_{60}$ doped with Kr show evidence that this orientational glass undergoes polymorphous transformation and two orientationally disordered phases coexist in the interval 2-24 K. The phenomenon of polymorphism has been known for the case of structural glasses though it is not clear completely yet. As for orientational glasses, it seems to be found here for the first time. Below we use the term "polyamorphism" which was introduced in analogy with "polymorphism" in [34] and then was reintroduced by others [35]. Note that when the sample is heated from 2 K to 24 K and cooled again to 2 K, its volume decreases by ΔV=0.017% because of the thermal expansion hysteresis. ΔV can be treated as the lower limit for the difference between the molar volumes of the glasses mentioned.
In the future we are planning more detailed investigations of polyamorphism in this type of orientational glasses.

At the lowest temperatures of the experiment, the glass with a larger molar volume is more advantageous thermodynamically. The thermal expansion of this glass is determined by two contributions (see Fig.1.). The lattice excitations (phonons and librons) make a positive contribution to the thermal expansion, while the molecule reorientations contribute negatively. Earlier, we attributed the negative thermal expansion of pure $C_{60}$ to the tunnel reorientation of its molecules. The Kr atoms occupying the octahedral interstitial sites of $C_{60}$ increase the volume of the crystal and the degree of disordering in it. This should suppress the barriers impeding the rotation and increase the probability of rotational tunneling. As a result, we can expect that, the negative contribution to the thermal expansion would increase and the temperature interval, where this contribution exists, would become wider. This is what we observe experimentally.

It is interesting that the difference between the α(T) curves taken on heating and cooling (Fig.6) coincides, within the experimental error, with the value of the negative contribution to the thermal expansion. Thus, the contributions of the lattice excitations to the thermal expansion of the two types of glasses are practically equal. The difference in the thermal expansion of these glasses is due to the processes molecular reorientation by $C_{60}$.

The increase in the molar volume of $C_{60}$ on its doping with Kr should suppress to some extent the frequencies of translational and orientational vibrations of the $C_{60}$ lattice. As a result, the contribution of the lattice excitations to the thermal expansion of $C_{60}$ should be larger in the studied interval of temperatures. We believe that this is the reason why the thermal expansion coefficient measured on cooling $C_{60}$(IV)+Kr exceeds that of the initial $C_{60}$(IV) sample.

The introduced (Eq. (2)) relaxation time $\tau_1$ and $\tau_2$ describe the temperature equalizing over the sample (thermalization) and the orientational relaxation, respectively. As seen in Fig. 8, the process of orientational relaxation is slower. Commonly, in thermally activated processes $\tau_2$ does not increase with temperature. It is another argument in favor of tunnel reorientation in a part of $C_{60}$ molecules.

**3.3.2 $C_{60}$ – He.** The starting sample was $C_{60}$(III). It was previously used in different experiments and was therefore saturated with Ar, Ne and $D_2$. Employing prolonged evacuation of the sample at room temperature, we succeeded in removing Ar and Ne nearly completely. The return of the thermal expansion coefficients of the previously doped sample to the values of pure $C_{60}$ (III) was taken as an indication that desaturation was completed. As for $D_2$, keeping $C_{60}$-$D_2$ in vacuum at room temperature for 180 days was not sufficient to remove all dissolved gas, and we had to prolong the procedure for 48 hours more at T=250 C. But even after this vacuum exposure the desaturation was not complete. The thermal expansion of the sample was still positive at liquid helium temperature. In the investigated interval of temperatures the thermal expansion of this sample practically coincided with that of $C_{60}$(IV) and no hysteresis was detected. To saturate it with He, the sample was kept in the dilatometric cell for 24 hours at room temperature under the He pressure of 1 atm. The dilatometric cell was then cooled for 7 hours to T=4.2 K. During cooling the He pressure decreased in the closed volume of the dilatometer, and helium was added at 112 K and 45 K to restore the pressure to 1 atm.

At 4.2K the He gas was evacuated from the dilatometer and, while measuring the thermal expansion, the sample $C_{60}$-He was kept in vacuum. No indications of the sample desaturation were observed during dilatometric measurements.

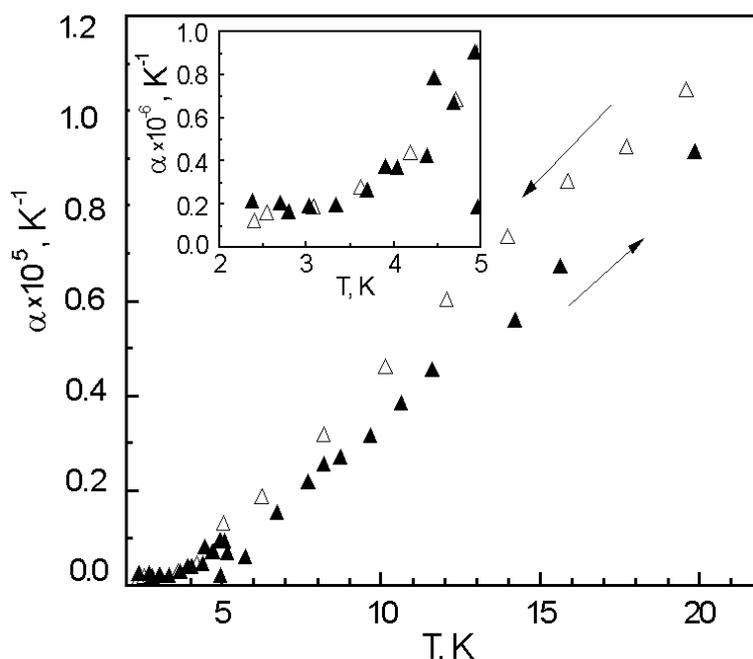

**Fig. 9.** Temperature dependence of the thermal expansion coefficients of He-doped fullerite

The temperature dependence of the linear expansion coefficient of $C_{60}$+He is shown in Fig.9 for the heating (solid triangles) and cooling (open triangles) conditions. Along with $C_{60}$(IV)+Kr, the thermal expansion of $C_{60}$+He has some peculiarities, such as (a) the hysteresis, (b) the negative contribution to the thermal expansion observed on heating and (c) the $\alpha$ - values measured on cooling which exceed the coefficients for the initial $C_{60}$ sample. In contrast to $C_{60}$(IV)+Kr, for $C_{60}$+He the hysteresis is smaller in magnitude and appears only if the considered temperature range lies greater than 5 K. The negative contribution to the thermal expansion of $C_{60}$+He is always smaller in absolute value than the positive one. The heating of the sample to the highest temperature (20 K) and the subsequent cooling to 2 K lead to a $4.6 \times 10^{-3}$% decrease in the volume of the sample. For $C_{60}$(IV)+Kr the decrease is $17 \times 10^{-3}$%.

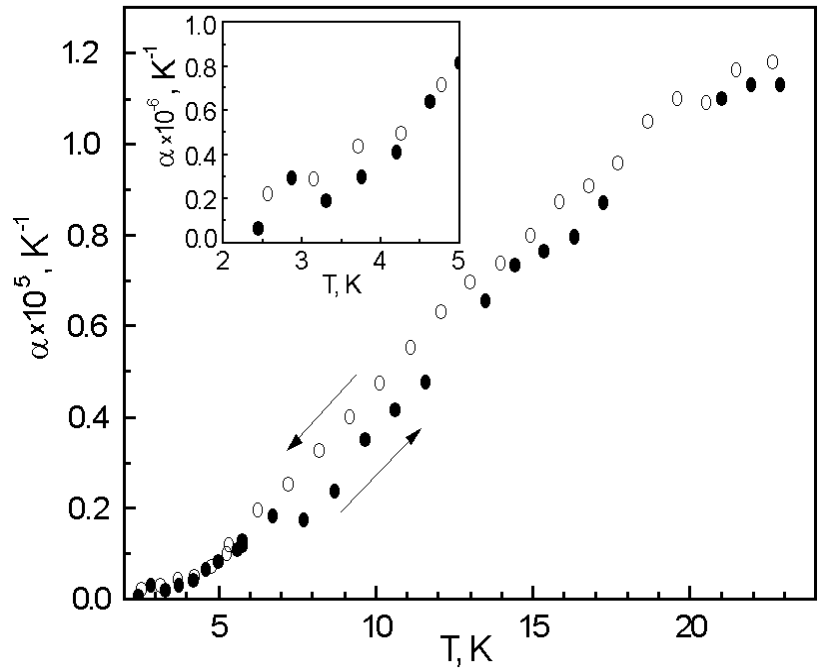

**Fig. 10.** Temperature dependence of the thermal expansion coefficients of He-doped fullerite after partial desaturation.

Next the $C_{60}$+He sample was partially desaturated. To do this, the sample was heated to 250 K for 11 hours during continuous dynamic evacuation and then cooled to 4.2 K for 7 hours. The thermal expansion coefficients of $C_{60}$+He measured after the partial desaturation are shown in Fig.10 (solid circles for heating and open circles for cooling). After the partial desaturation, the hysteresis became narrower, and the relative change in the volume caused by heating the sample to 23K and subsequent cooling to 2K decreased to $3.2\times10^{-3}$%.

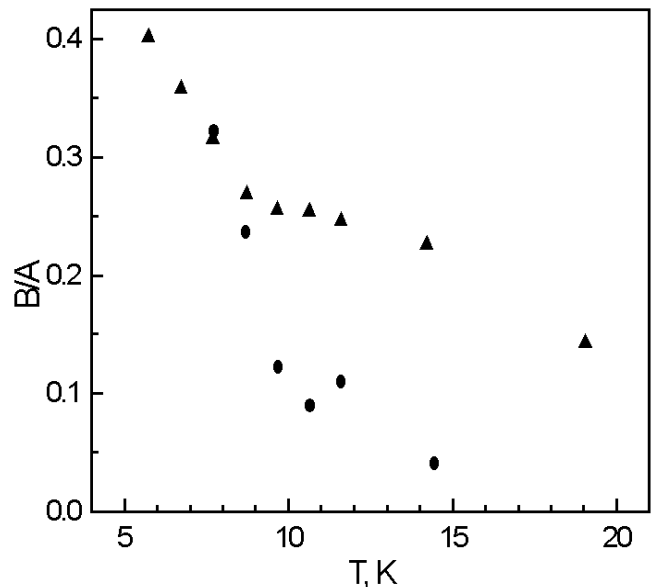

**Fig. 11.** The ratio between the absolute values of the negative and positive contributions to the thermal expansion of $C_{60}$ doped with He:
▲ – before partial desaturation;
• – after partial desaturation.

The temperature dependence of the absolute value ratio between the negative and positive contributions to the thermal expansion coefficient measured on heating the sample is shown in Fig.11. As we could expect, after the partial desaturation the negative contribution, i.e. the contribution made by reorientation of the $C_{60}$ molecules, becomes weaker.

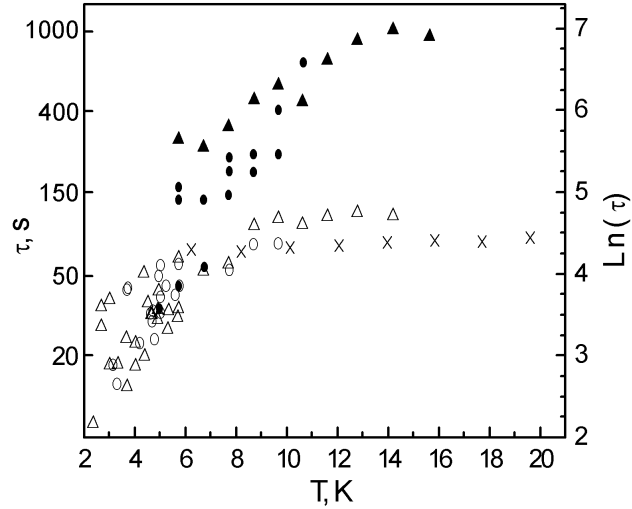

**Fig. 12.** Characteristic relaxation times $\tau$ of the thermal expansion of $C_{60}$+He:
$\Delta, \times$ - $\tau_1$; $\circ$ - $\tau_2$ – before partial desaturation;
$\blacktriangle$ - $\tau_1$;, $\bullet$ - $\tau_2$ – after partial desaturation.

The temperature dependences of the relaxation times of the positive ($\tau_1$) and negative ($\tau_2$) contributions to the thermal expansion of $C_{60}$ doped with He are shown in Fig.12. It is seen that the partial desaturation does not produce much influence on $\tau_1$ and $\tau_2$. We have found no qualitative distinctions between the temperature dependences of the relaxation times of the $C_{60}$+He and $C_{60}$(IV)+Kr solutions. At T>12K the orientational relaxation times $\tau_2$ are larger for $C_{60}$+He than for $C_{60}$(IV)+Kr.

The detected features of the thermal expansion of fullerite $C_{60}$ doped with helium can be explained in the framework of the model that we proposed for $C_{60}$+Kr.

**4 Conclusions**

The low temperature thermal expansion of the orientational glass of pure $C_{60}$ and $C_{60}$ doped with inert gases is determined by the sum of positive and negative contributions. The analysis of their magnitudes, temperature dependences and the related relaxation times has permitted us to assume that the positive contribution is made by the lattice excitations (phonons and librations excitations), while the negative contribution results from tunneling reorientation of a part of $C_{60}$ molecules.

The implantation of Ar and Ne impurities decreasing the molar volume of fullerite enhances the noncentral interaction between the $C_{60}$ molecules. The change caused by the impurities in the thermal expansion of $C_{60}$ was interpreted as evidence of lower probability of rotational tunneling and higher libration frequency of the $C_{60}$ molecules in the lattice of fullerite.

The implantation of Kr and He impurities increasing the molar volume of $C_{60}$ makes the noncentral interaction between the $C_{60}$ molecules weaker. This generates two effects. First, the probability of rotational tunneling of the $C_{60}$ molecules becomes higher; hence, the contribution of rotational tunneling to thermal expansion increases and the temperature interval, where this contribution exists, extends. Second, there is hysteresis of the thermal expansion of doped fullerite. The hysteresis may be due to the formation and mutual conversion of two phases of this orientational glass, which permits us to extend the phenomenon of polyamorphism on orientational glasses as well.